\documentclass[conference]{IEEEtran}
\IEEEoverridecommandlockouts
\usepackage{cite}
\usepackage{amsmath,amssymb,amsfonts}
\usepackage{algorithmic}
\usepackage{graphicx}
\usepackage{textcomp}
\usepackage{xcolor}
\def\BibTeX{{\rm B\kern-.05em{\sc i\kern-.025em b}\kern-.08em
    T\kern-.1667em\lower.7ex\hbox{E}\kern-.125emX}}
\begin{document}

\title{Optimizing microservices with hyperparameter optimization}

\author{\IEEEauthorblockN{Hai Dinh-Tuan}
\IEEEauthorblockA{\textit{Service-centric Networking} \\
\textit{Technische Universit\"at Berlin}\\
Berlin, Germany \\
hai.dinhtuan@tu-berlin.de}
\and
\IEEEauthorblockN{Katerina Katsarou}
\IEEEauthorblockA{\textit{Service-centric Networking} \\
\textit{Technische Universit\"at Berlin}\\
Berlin, Germany \\
a.katsarou@tu-berlin.de}
\and
\IEEEauthorblockN{Patrick Herbke}
\IEEEauthorblockA{\textit{Service-centric Networking} \\
\textit{Technische Universit\"at Berlin}\\
Berlin, Germany \\
p.herbke@tu-berlin.de}
}

\maketitle

\begin{abstract}

In the last few years, the cloudification of applications requires new concepts and techniques to fully reap the benefits of the new computing paradigm. Among them, the microservices architectural style, which is inspired by service-oriented architectures, has gained attention from both industry and academia. However, decomposing a monolith into multiple microservices also creates several challenges across the application's lifecycle. In this work, we focus on the operation aspect of microservices, and present our novel proposal to enable self-optimizing microservices systems based on grid search and random search techniques. The initial results show our approach is able to optimize the latency performance of microservices to up to 10.56\%.

\end{abstract}


\section{Introduction}

In essence, microservices encourage a more agile and modular approach to the whole software lifecycle, from design, implementation, operation, to maintenance. By decomposing monoliths into smaller, independently deployable units, microservices can achieve a high level of scalability and resiliency \cite{dinh2019maia}. 

However, this design also comes at a cost: a microservices-based application comprises many distributed services, thus requires enormous efforts to manage individual services. In addition, a good optimization technique also needs to consider the complex interactions among microservices while tuning interference parameters.

The increasing complexity of software systems in general and cloud-based applications in particular gradually make manual optimization impracticable. Motivated by that, this work aims to automate the performance optimization process using techniques known from machine learning, especially in hyperparameter optimization problems.

\section{Use case}

\begin{figure}
	\includegraphics[width=\linewidth]{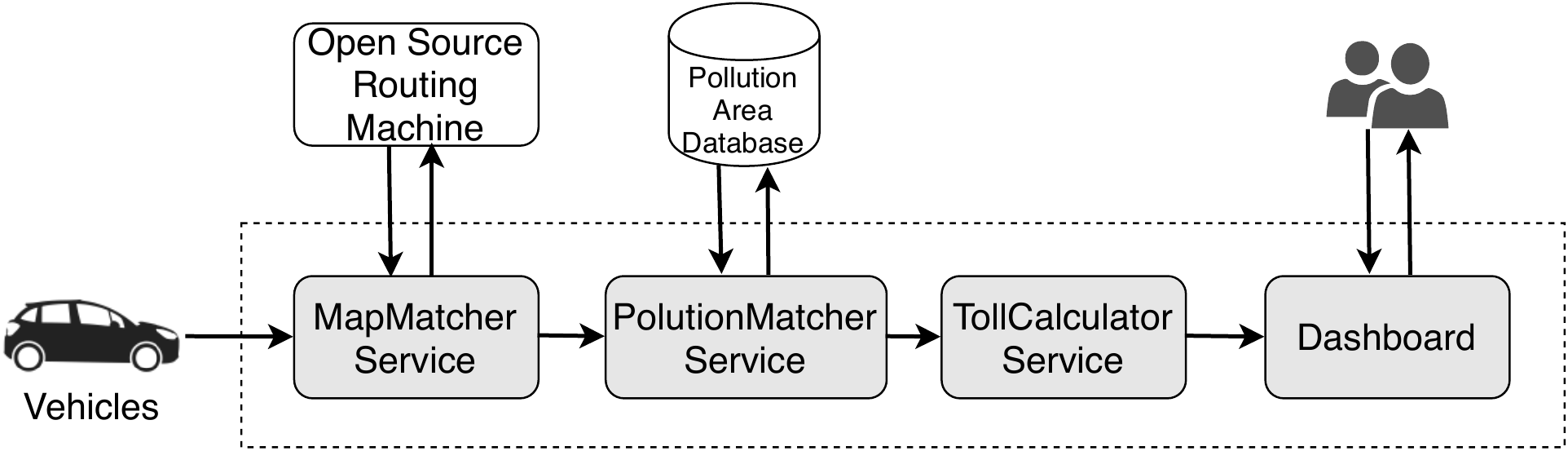}
	\caption{Overview processing flow of the air pollution-aware toll system.}
	\label{fig:overview}
\end{figure}

We evaluate our approach using an existing microservices-based application, which can automatically calculate the environmental toll for vehicles based on the pollution level \cite{dinh2020development}. The location coordinates of each vehicle are fetched in real-time into the application, where they will be processed by a chain of microservices as depicted in Figure \ref{fig:overview}. To evaluate the end-to-end latency performance as well as the latency performance by each microservice, we apply a technique called \textit{Distributed Tracing} \cite{shkuro2019mastering}.

\section{Background \& Concept}

Optimizing a microservices-based application with complex interference among various parameters has been identified as an exceptionally complex and challenging problem \cite{somashekar2021towards}. Several works have focused on employing new techniques to automate the optimization process, such as Bayesian optimization \cite{ramakrishna2019}, Meta-heuristic optimization \cite{thakkar2021spread}. 

In this work, we proposed using \textit{Grid search} and \textit{Random search}, which have been used extensively for optimizing hyperparameters in machine learning, to optimize the microservices' configuration during runtime. These two approaches require a bounded \textit{search space}, which includes possible values for each parameter. While Grid search evaluates every position in the grid, Random search only randomly evaluates parameter combinations.

In this work, we develop a new software component called \textit{Microrservice optimizer}, which automatically iterates through the parameters selected by the operator (the search space), generates new configuration combinations, applies them to the microservices, and observes the performance of the application as a whole. These performance data is stored after each iteration to find an optimal configuration setting for each individual microservice.

\begin{figure}
	\includegraphics[width=\linewidth]{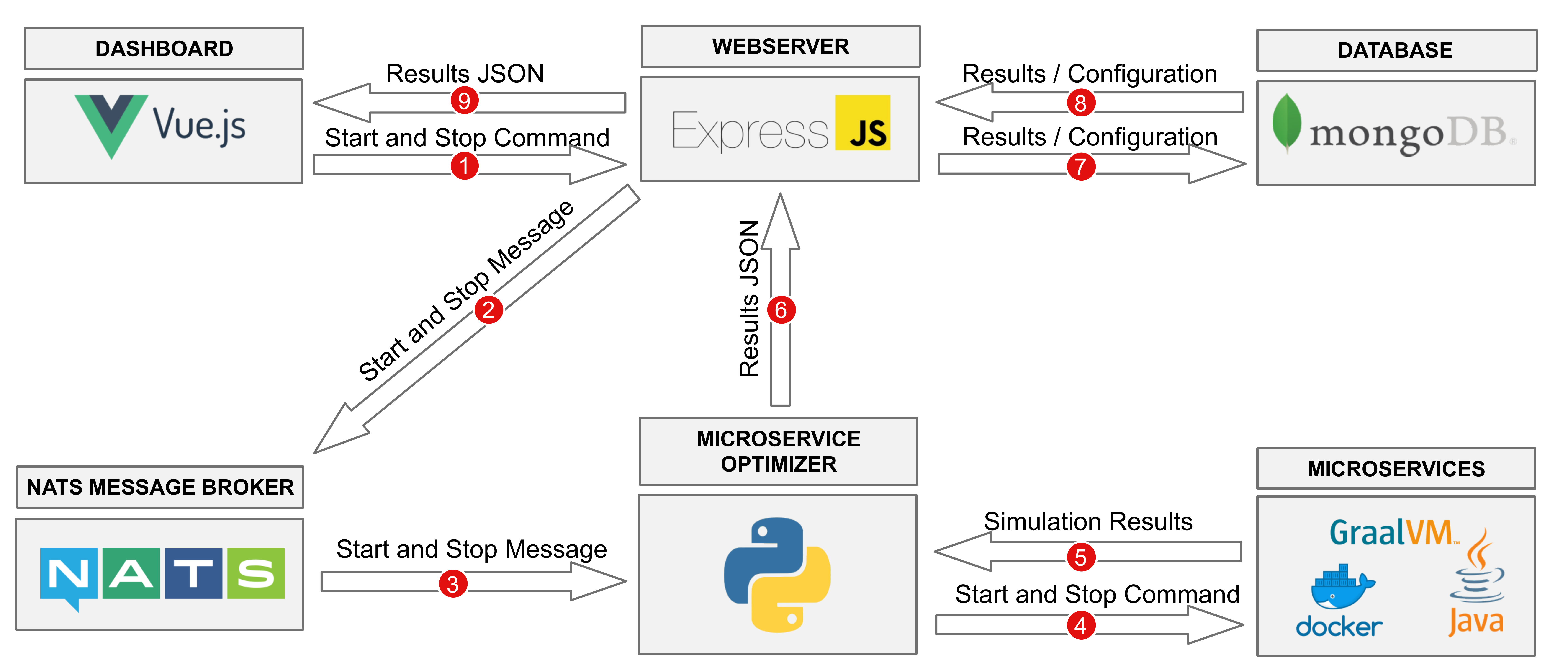}
	\caption{Architecture overview.}
	\label{fig:architecture}
\end{figure}

\begin{figure}
	\includegraphics[width=\linewidth]{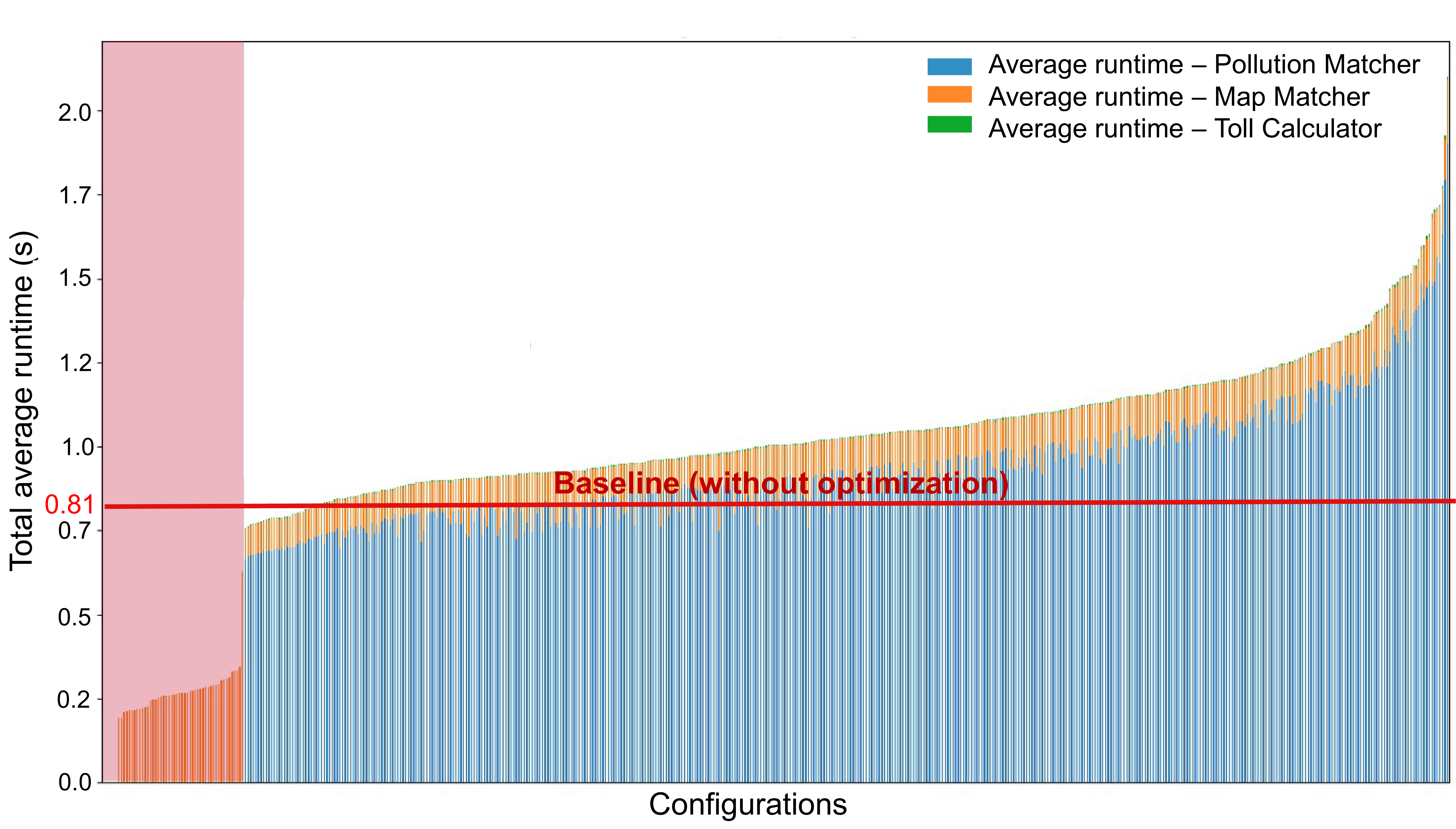}
	\caption{Evaluation results of Grid search optimization.}
	\label{fig:grid}
\end{figure}

\begin{figure}
	\includegraphics[width=\linewidth]{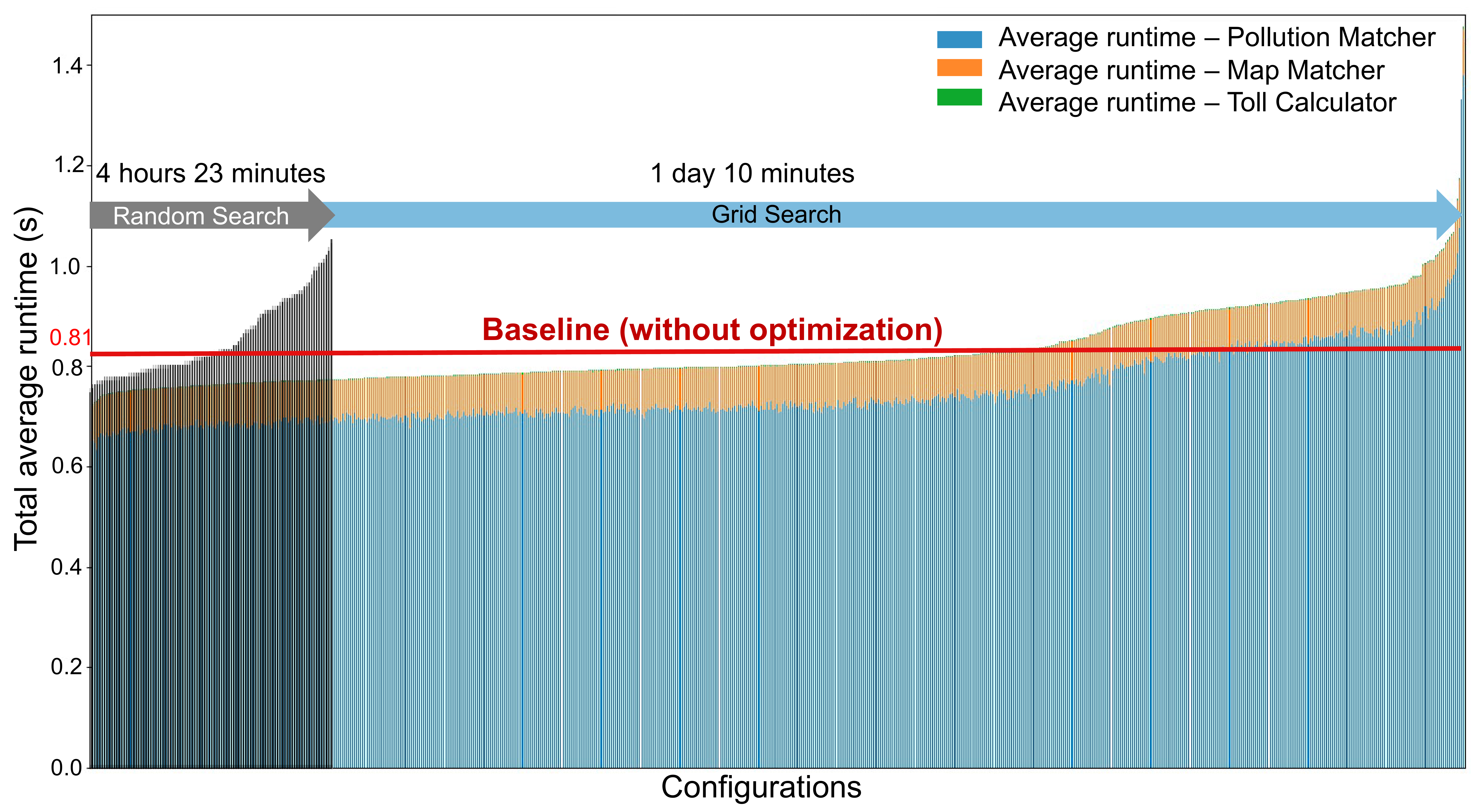}
	\caption{Evaluation results of Random search in compared with Grid search.}
	\label{fig:random}
\end{figure}

\section{Initial evaluation results}

All the microservices are developed in Java and deployed as Docker containers in an Amazon AWS EC2 t2.medium instance with 2vCPUs and 4GB of RAM. The communication among microservices are done by exchanging messages via a NATS broker instance. The test setup consists of six components as shown in Figure \ref{fig:architecture}: Dashboard (web application), Web Server, Database, NATS message broker, Microservice Optimizer and Microservices. The arrows and numbers explains the optimization workflow and how the components communicate with each other. The Dashboard allows operators to select a set of parameters to be included in the optimization process. In our prototype, tunable parameters include Java virtual machine (JVM or GraalVM) and Docker parameters, which can be further categorized into four main types: Boolean, Discrete, Byte, and Categorical parameters. With an initial set of 11 parameters, the search space has in total 177,147 configuration combinations. The Web server together with the Microservice Optimizer coordinate the iteration through the search space defined by the operator. The Database stores evaluation results, while the NATS message broker is used by both internal communication among microservices and the communication between the Web Server and the Microservice Optimizer.

Figure \ref{fig:grid} shows the average runtime per configuration while using Grid search, sorted by latency. Please note that the red area marks incomplete runs. The latency of a non-optimized baseline run is marked with a red line with an average end-to-end latency of 0.8100s. The best performing configuration identified by Grid search has an average latency of 0.7245, yield 10.5556\% improvements. 

Figure \ref{fig:random} shows the average runtime per configuration in Random search (black) and Gird search (colored), sorted by latency. The evaluation results show that, the best performing configuration identified by Random search has an average latency of 0.7445, yield 8.0864\% improvements, which is not as optimal as the best performing configuration identified by Grid search. However, using Random search can find a near-optimal solution 84\% time faster than Grid search approach.

\section{Conclusions}

This short paper presents our immediate results of a new microservice optimizer based on two methods, which are Grid search and Random search. Our first prototype focuses on optimizing Java virtual machines and Docker container parameters; therefore, this approach can be applied to every microservice-based application that uses these technologies.

Our evaluation shows that our microservice optimizer can find configurations that reach a notable improvement over a default, non-optimized set of parameters. In addition, when comparing the two methods, the Random search-based approach can identify a comparable configuration in terms of latency performance while reduce the calculation time significantly. In the next phases, we will expand the parameter search space and investigate other optimization strategies.

\section*{Acknowledgment}
We are grateful for the dedicated contributions to this project by Jonathan Kossick and Adrian Michalke.

\bibliographystyle{IEEEtran}
\bibliography{IEEEabrv,references.bib}

\end{document}